\documentclass{JHEP3}

\usepackage{amsmath}
\usepackage{amssymb}

\def\setC{\mathbb{C}}

\def\setR{\mathbb{R}}

\newcommand{\dd}{\mathrm{d}}

\newcommand{\ie}{\textsl{i.e.~}}

\newcommand{\eg}{\textsl{e.g.~}}
\newcommand{\etal}{\textsl{et al.~}}

\newcommand{\bx}{\mathbf{x}}

\newcommand{\GN}{G_{_{\rm N}}}

\newcommand{\GReCO}{${\cal G}\setR\varepsilon\setC{\cal O}$}

\newcommand{\Phir}{\Phi{_\mathrm{r}}}
\newcommand{\Ps}{\Phi{_\mathrm{s}}}
\newcommand{\Hu}{{\cal H}} \newcommand{\Ka}{{\cal K}}
 \newcommand{\cs}{{c_\mathrm{_S}^2}}

\newcommand{\cS}{c_{_\mathrm{S}}}
\newcommand{\cZ}{c_{_\mathrm{Z}}}

\newcommand{\lP}{\ell_{_\mathrm{P}}}
\def\wun{\omega_1}
\def\wdeu{\omega_2}
\def\eun{\epsilon_1}
\def\edeu{\epsilon_2}
\def\xnec{x_{_\mathrm{NEC}}}

\title{Adiabatic and entropy perturbations propagation in a bouncing
universe.}

\author{Patrick Peter\\ Institut d'Astrophysique de Paris -- \GReCO,
\\ FRE 2435-CNRS, 98bis boulevard Arago, \\ 75014 Paris, France. \\ E-mail:
\email{peter@iap.fr} }

\author{Nelson Pinto-Neto\\ Centro Brasileiro de Pesquisas F\'\i
sicas, \\ Rue Dr.  Xavier Sigaud 150, Urca 22290-180 \\ Rio de
Janeiro, RJ, Brazil. \\ E-mail: \email{nelsonpn@cbpf.br}}

\author{Diego A. Gonzalez\\ Centro Brasileiro de Pesquisas F\'\i
sicas, \\ Rue Dr.  Xavier Sigaud 150, Urca 22290-180 \\ Rio de
Janeiro, RJ, Brazil.\\ E-mail: \email{diego@cbpf.br}}

\preprint{September 8$^\mathrm{th}$, 2003}

\abstract{By studying some bouncing universe models dominated by a
specific class of hydrodynamical fluids, we show that the primordial
cosmological perturbations may propagate smoothly through a general
relativistic bounce. We also find that the purely adiabatic modes,
although almost always fruitfully investigated in all other contexts
in cosmology, are meaningless in the bounce or null energy condition
(NEC) violation cases since the entropy modes can never be neglected
in these situations: the adiabatic modes exhibit a fake divergence
that is compensated in the total Bardeen gravitational potential by
inclusion of the entropy perturbations.}

\begin{document}

\section{Introduction}

The singularity problem in cosmology~\cite{singularity}, which is not
addressed in the framework of inflationary paradigm~\cite{inflation},
(note however the recent ``emergent universe''
proposal~\cite{emergent} in that respect) may be solved either through
a Pre-Big-Bang (PBB) type phase~\cite{PBB} or by means of a
bounce~\cite{tolman,seventies}. Both cases in fact demand that the
universe undergoes at least one transition from a collapsing to an
expanding epoch, although the PBB situation demands that this happens
in the Einstein frame in which general relativity (GR) holds true. It
is this kind of bounce that we shall be concerned with here.

A bouncing phase may originate from quantum gravity~\cite{QG} or
quantum cosmology~\cite{QC}, and have been seen to occur in some
string-motivated models~\cite{FFPP}. In one such model, based on the
brane hypothesis in a five dimensional context, namely the ekpyrotic
scenario~\cite{ekp},\cite{noekp}, the bounce in fact does not address
the singularity problem since it is assumed to be singular.  This
leads to some difficulties~\cite{noekp} which the model still have to
deal with (see Ref.~\cite{ekp5D}).

It is worth pointing out that in some improvements of the regular
bouncing models, not necessarily relying on general relativity at all
times, perturbations may even be made scale-invariant~\cite{ppnpn2},
thus moving one step towards an alternative model of the Universe,
namely one for which there is only a {\sl short} period of accelerated
expansion, which happens during the bouncing era following a
contraction phase, in contrast with the relatively {\sl long} period
characteristic of usual inflationary scenarios.

Bouncing models without a long period of inflation afterwards, besides
solving the singularity and the horizon problems, can be made to avoid
the trans-Planckian issue~\cite{transP} and may also provide a causal
explanation, in terms of a quantum mechanical origin, for the large
scale structures in our Universe. However, they yet depend on some
theory of initial conditions coming from quantum cosmology and/or
string theory in order to address the flatness and isotropization
issues, which are solved naturally in usual inflationary scenarios. In
short, bouncing models can somehow be viewed either as potential
alternatives to inflation, or as complementary to it, providing help
in constructing a full consistent cosmological model for our
Universe~\cite{PBB,tolman,seventies,ekp} and addressing issues which
inflationary models do not.

In the present work, we are interested in pointing out the relevance
of entropy fluctuations during the bounce phase, in particular at the
time at which the transition from an epoch where the null energy
condition (NEC) is valid to an epoch where it ceases to be valid
occurs (which will be called, from now on, the NEC transition), which
may affect the observable power spectrum of adiabatic perturbations in
the present expanding phase of the abovementioned bouncing models,
including the PBB case.

In a previous study~\cite{nobounce}, concentrating on pure general
relativity as the theory for describing gravity, we showed that a
bouncing universe described in terms of hydrodynamical fluids was
unstable with respect to adiabatic perturbations. More
recently~\cite{ppnpn2}, we managed to obtain a model in which all
perturbations are finite at all times, including the full Bardeen
potential without imposing any adiabatic condition, by considering the
case of a two fluids bouncing Universe, one of the fluids described by
a scalar field.

The case of Ref.~\cite{ppnpn2} (see also Ref.~\cite{NS}) is but one
particular case of Ref.~\cite{nobounce}, called {\it case 4} there,
with the scale factor behaving near the bounce as an even function of
the conformal time $\eta$, at least up to the fifth order. The bounce
itself, \ie the point at which the scale factor derivative with
respect to time vanishes, is stable in this case, in the sense that
the perturbations are bounded while passing through it.  Instabilities
appear however as logarithmic divergences in the second derivative of
the adiabatic part of the gauge invariant Bardeen potential, hence
also in the Einstein tensor and pressure perturbations, at the NEC
transition point. In fact, any model presenting a NEC transition faces
the abovementioned divergence in the adiabatic Bardeen potential.

The aim of this paper is to show, in a general framework, that the
instability of adiabatic perturbations at the NEC transition point
does not stem from a possible ill-defined decomposition of the
pressure perturbation into adiabatic and entropy parts, but resides on
the fact that entropy fluctuations cannot be neglected there, even for
arbitrarily large wavelengths. Note that in the framework of the
present article, entropy modes mean modes of mixing entropy between
the various fluid constituents of the universe, as opposed to
intrinsic entropy modes, which we assume are absent. Hence, any
attempt to calculate the power spectrum of perturbations on bouncing
models relying only on matching conditions for the adiabatic Bardeen
potential through the bounce is not sufficient and may lead to
erroneous results. The entropy fluctuations, in particular at the NEC
transition point, must be considered because they are not negligible,
in fact they are extremely important for the stability of the model,
and they may transfer power to the adiabatic perturbations
afterwards. We also settle down general conditions in which
perturbations can be consistently defined in a bouncing model.

\section{Bouncing Background}\label{sec:back}

Let us begin by a review of the main general aspects of bouncing
backgrounds. Within our conventions, the FLRW metric reads
\begin{equation} \dd s^2 = a^2(\eta) \left( \dd\eta^2 - \gamma_{ij}
\dd x^i \dd x^j \right),
\label{FRW}\end{equation}
with the spatial three-metric $\gamma_{ij}$ given by
\begin{equation}
\gamma_{ij} \equiv \left(
1+\frac{\Ka}{4} \bx^2 \right)^{-2} \delta_{ij},
\label{3D}
\end{equation}
and $\eta$ being the conformal time from which one derives the cosmic
time $t$ as the solution of the equation $a \dd\eta = \dd t$, with a
given scale factor $a(\eta)$. In Eq.~(\ref{3D}), the parameter $\Ka$,
representing the spatial curvature, can be normalized to $\Ka=0,\pm
1$, and $\bx^2 \equiv \delta_{ij} x^i x^j$ (units are such that we are
setting $\hbar =c=1$).

The stress energy tensor, source of Einstein field equations, will
take the form
\begin{equation} T^\mu_\nu = (\epsilon + p)u^\mu u_\nu - p
\delta^\mu_\nu, \label{Tmunu}\end{equation} for energy density
$\epsilon$, pressure $p$ and 4-velocity (fluid tangent vector) {\bf
u}, which can be expressed as $u^\mu = (1/a) \delta^\mu_0$, \ie,
$u_\nu = a \delta_{\nu 0}$. Einstein equations for the metric
(\ref{FRW}) and stress energy tensor (\ref{Tmunu}) read
\begin{equation}
\Hu^2 + \Ka = \lP^2 a^2 \epsilon,\label{EE01}
\end{equation}
and
\begin{equation}
\beta \equiv \Hu^2 -\Hu' + \Ka = \frac{3}{2} \lP^2 a^2 (\epsilon + p),
\label{EE02}
\end{equation}
where $\lP \equiv 8\pi \GN/3$ is the Planck length ($\GN$ being Newton
constant), $\Hu\equiv a'/a$ the conformal Hubble parameter, and a
prime denotes a derivative with respect to the conformal time
$\eta$. At the point of NEC transition we have $(\epsilon + p)=0$, \ie
$\beta=0$.

We now show that in the case where the energy density can be written
as a function of the scale factor $a(\eta)$, one can prove that this
function is even around the bounce, behaving as
\begin{equation}
\label{joao}
a(\eta)=a_0 + b\eta ^2 + e\eta ^4 + \cdots,
\end{equation}
with $\eta=0$ representing the bounce. Note that such a bounce needs a
NEC transition point\footnote{In fact, a NEC transition point may not
be present in such models if the curvature of the spatial sections is
positive and if $2b\leq a_0$. However, in order for a realistic model
to satisfy these conditions, one needs either to impose a tremendous
amount of fine-tuning, or to have an inflationary phase between the
bounce and the radiation dominated phase; see Ref.~\cite{nobounce} for
details.}. Indeed, the scale factor $a(\eta)$ of a general bouncing
model can be expanded around the bounce as follows:
\begin{equation}
\label{abounce}
a=a_0 + b\eta ^{2n} + d\eta ^{2n+1} + e\eta ^{2n+2} + f\eta ^{2n+3} +
\cdots\; ,
\end{equation}
where $a_0 >0$, and the integer $n$ satisfies $n\geq 1$. This means we
demand that $a(\eta)$ must be at least $\mathrm{C}^{2n+3}$ near the
bounce. In order that Eq.~(\ref{abounce}) indeed represents a bounce,
the otherwise arbitrary parameter $b$ must satisfy $b>0$.

The function $\Hu (\eta)$ coming from Eq.~(\ref{abounce}) reads
\begin{eqnarray}
\label{Hbounce}
\Hu=\frac{1}{a_0^2} &\big[& 2nba_0\eta ^{2n-1} + (2n+1)d a_0
\eta^{2n}+2(n+1) e a_0\eta ^{2n+1}+(2n+3)fa_0\eta
^{2n+2} \nonumber \\ & & - 2nb^2 \eta^{4n-1} - bd(4n+1)
\eta^{4n}+\cdots \big],
\end{eqnarray}
while $\beta(\eta)$ is
\begin{eqnarray}
\label{bbounce}
\beta&=&\frac{1}{a_0^2}\big[ \Ka a_0^2 - 2n(2n-1)ba_0\eta ^{2n-2}
 - 2n(2n+1)da_0 \eta ^{2n-1} - 2(n+1)(2n+1)ea_0\eta
^{2n}\nonumber\\ & & - 2(2n+3)(n+1)fa_0 \eta ^{2n+1} + 2n(6n-1)b^2\eta
^{4n-2}+ 8bdn(1+3n) \eta^{4n-1}+\cdots\big ].
\end{eqnarray}
In the case where the energy density can be written as a function of
$a$, $\epsilon_0 = \epsilon_0(a)$, one can prove that $a(\eta)$ is
even and that $n=1$ in Eq.~(\ref{abounce}). This can be seen as
follows: from the energy-momentum conservation equation we obtain
$a\dd\epsilon/\dd a =-3(\epsilon + p)$, implying that $p$ is also a
function of $a$, $p=p(a)$. As a consequence, the right-hand-side of
Eq.~(\ref{EE02}) implies that $\beta$ can also be written as a
function of $a$, and hence it can be Taylor expanded as
\begin{equation}
\beta(a)\simeq  \beta(a_0)+ \beta_{,a} (a_0)\left( b\eta ^{2n} +
d\eta ^{2n+1} + \cdots \right)+
\frac{\beta_{,aa}(a_0)}{2} \left(b\eta ^{2n}+ d\eta ^{2n+1} +
\cdots\right )^2, \label{expandf}
\end{equation}
where the ``$,a$'' indices indicate differentiation with respect to
the scale factor. Equations (\ref{bbounce}) and (\ref{expandf}) are,
respectively, the left hand side (LHS) and RHS of
Eq.~(\ref{EE02}). Comparing powers of the conformal time, and
identifying the coefficients, one finds that the term containing
$d\eta ^{2n-1}$ in the LHS has no counterpart in the RHS. Hence, $d$
must vanish. With $d=0$, the term containing $f\eta ^{2n+1}$ in
Eq.~(\ref{bbounce}) also has no corresponding term in
Eq.~(\ref{expandf}), so that $f$ must vanish as well. By induction, we
can prove that all coefficients of terms containing odd powers of
$\eta$ in the expansion Eq.~(\ref{abounce}) must be zero. Hence,
$a(\eta)$ must be an even function of the conformal time.\footnote{As
an alternative argument, let us note that $\epsilon (\eta) = \epsilon
[\eta(a)] = \epsilon (a)$ is true if and only if the inverse of
$a(\eta)$, $\eta(a)$, exists, which is possible, in a bounce framework
(\ie if $a$ is not a monotonic function of $\eta$), if and only if
$a(\eta)$ is even, as $a$ is positive.} Also, if $n>1$, the term
containing $\eta ^{2n-2}$ in Eq.~(\ref{bbounce}) has no counterpart in
Eq.~(\ref{expandf}); this imposes that $n=1$, and hence a scale factor
that behaves quadratically in $\eta$.

A realistic bouncing universe, \ie one leading smoothly to a radiation
dominated phase, cannot be modeled by means of a single barotropic
fluid with constant equation of state, since this would imply the
single fluid in question to consist in radiation, and it is well-known
that radiation alone cannot prevent the occurrence of a singularity
forming~\cite{singularity}, even for positively curved spatial
sections.  One could also try to describe such a bounce with just one
fluid by allowing a varying equation of state for this fluid, assuming
the equation of state parameter $\omega\equiv p/\epsilon$ (with
$\epsilon$ the energy density and $p$ the pressure) to depend on time
in such a way that $\lim \omega =1/3$ for large times.  This case,
somehow implicitly assumed in Ref.~\cite{nobounce}, will not be
treated here since it would demand complete knowledge of the behavior
of the intrinsic entropy of the fluid, which is model-dependent. We
shall accordingly in what follows consider the next-to-simple case of
$N$ constant equations of state components, later allowing for
exchanges in the various modes of perturbations they may produce at a
given scale, thereby providing the possibility to introduce mixing
entropy modes in a model-independent way.

We want to concentrate on the situation relevant for the rest of the
paper for which the matter content is described by an arbitrary number
of non interacting hydrodynamical perfect fluids, each with a
barotropic equation of state with constant ratio between the energy
density and pressure; this is but a special case of the one where the
total energy density can be written as a function of the scale factor,
and hence with even scale factor, bouncing behavior of
Eq~(\ref{joao}), and a NEC transition point. This means that the
components are noninteracting perfect fluids with stress-energy
tensors given by
\begin{equation}
T_\mathrm{tot} ^{\mu\nu} = \sum_{i=1}^N T_i^{\mu\nu} , \qquad T_i
^{\mu\nu} = (\epsilon_i + p_i) u^\mu u^\nu - p_i g^{\mu\nu},
\label{Tmunui}
\end{equation}
relations in which $u_\mu u^\mu =1$ is the same timelike vector for
all fluids, in agreement with the symmetry assumptions leading to a
FLRW Universe. In the simple case with which we are interested here,
we demand that the fluid equations of state be fixed, namely
\begin{equation}
p_i = \omega_i \epsilon_i, \qquad \omega_i = \hbox{const.}
\end{equation}
The total energy density and pressure that enter Einstein equations
are $\epsilon = \sum_i \epsilon_i$ and $p=\sum_i p_i$.
Energy-momentum conservation for each fluid, $\nabla_\mu
T_i^{\mu\nu}=0$ implies
\begin{equation}
\epsilon_i' + 3\Hu \epsilon_i (1+\omega_i) = 0 \qquad \Longrightarrow
\qquad \epsilon_i = c_i a^{-3 (1+\omega_i)},
\end{equation}
with $c_i$ arbitrary constants. Note at this point that the energy
density in this case is indeed a function of $a$, so that, from the
arguments above, $a(\eta)$ must be even with $n=1$ in the expansion
(\ref{abounce}). This situation corresponds to the so-called {\it case
(4)} of Ref.~\cite{nobounce}, which we want to investigate in greater
details below.

 Specializing to two fluids and upon using Eq.~(\ref{EE01}), one gets
\begin{equation}
\left(\frac{a'}{a}\right)^2 = \lP^2 a^{-(1+3\wun)} \left[ c_1 + c_2
a^{3(\wun-\wdeu)}\right] - \Ka ,
\label{Fried}
\end{equation}
{}from which it is clear that for non-positive curvature spatial
section, \ie for $\Ka \leq 0$, it is only possible to have a bounce,
as a point in time for which $a'=0$, provided one of the fluids has
negative energy; this means that for $\Ka \leq 0$, one of the
constants $c_1$ or $c_2$ ought to be negative. The case with positive
spatial curvature can allow both constants to be positive, but none
can vanish if one demands radiation to be present, and therefore that
either $\wun$ or $\wdeu$ be equal to one third.

As an aside, let us remark that, at the background level, a scalar
field $\phi$, with dynamics stemming from the action
\begin{equation}
{\cal S}_\mathrm{s} = \int \left[ \frac{1}{2} \left( \nabla_\mu
\phi\right) \left( \nabla^\mu \phi \right) - V(\phi)
\right]\sqrt{-g}\dd^4 x,
\label{scalaction}
\end{equation}
is equivalent to a perfect fluid with varying equation of state since
one has
\begin{equation}
\label{anaflu}
p_\mathrm{s} = \frac{1}{2 a^2} ({\phi'})^2 - V(\phi), \qquad
\epsilon_\mathrm{s} = \frac{1}{2 a^2} ({\phi'})^2 + V(\phi).
\end{equation}
This is no longer true at the perturbation level, unless the potential
$V$ vanishes, which is then equivalent to a stiff matter fluid $\omega
=1$: when $V=0$, one has indeed $\phi '\propto 1/a^2 \rightarrow
\epsilon_{\phi} \propto 1/a^6$. If the kinetic terms in
Eq.~(\ref{anaflu}) are negative, and including also a radiation fluid,
one recovers the prototypical bouncing model already discussed in
Ref.~\cite{ppnpn2}, which we shall discuss later on in
Sec.~\ref{sec:pertspec}. In the case for which the Universe is
positively curved, a bounce is possible even if it is dominated by a
single scalar field~\cite{hawking,turok,jmpp}.

We now turn to the perturbations in this class of models, which we
wish to expand on either a basis of a fluid-by-fluid decomposition, or
into adiabatic and entropy modes.

\section{General perturbations.}\label{sec:genpert}

The metric for a perturbed universe with no anisotropic stress
contribution can be written, in full generality, in the longitudinal
gauge (we use the notations of Ref.~\cite{mfb})
\begin{equation}
\dd s^2 = a^2(\eta) \left[ (1+2 \Phi) \dd\eta^2 - (1-2\Phi)\gamma_{ij}
\dd x^i \dd x^j \right],
\label{dg}
\end{equation}
where $\Phi$ is the gauge invariant Bardeen potential~\cite{bardeen},
which is unique since we aim at describing a situation with no
anisotropic pressure. The perturbed Einstein equations with this
metric then read
\begin{eqnarray}
\nabla^2 \Phi - 3 \Hu \Phi' - 3 (\Hu^2 -\Ka) \Phi &=& 4 \pi G a^2
\delta \epsilon^{\mathrm{(gi)}}, \quad \label{EE11} \\ \Phi'' + 3 \Hu
\Phi' + (2 \Hu'+\Hu^2 -\Ka) \Phi &=& 4 \pi G a^2 \delta p^{{\rm
(gi)}}.\quad \label{EE12}
\end{eqnarray}
All these quantities are defined in such a way as to be gauge
invariant, as emphasized by the superscript ``(gi)'' (see
Ref.~\cite{mfb}). We shall for now on take it for granted that all
quantities under consideration are gauge invariant, and therefore
avoid the use of the notation ``(gi)'', which will be implicit.

As a first step, in the subsection below, we show that if the matter
content of a background bouncing model can be described by an
arbitrary number of non interacting fluids with constant equations of
state, then the full Bardeen potential and all its derivatives are
completely regular at all times. Then, in the following subsection, we
define adiabatic and entropy fluctuation modes. We recover the general
results of Ref.~\cite{nobounce} that there are always divergences in
the gauge invariant adiabatic potential around the bounce, except in
the case with bouncing behavior of Eq.~(\ref{joao}), and around the
NEC transition, but using a different method: instead of approximating
the equations near the singular points and solving them afterwards, we
adopt the more accurate and rigorous method of solving the exact
equation and then expanding the corresponding solution and its
derivatives near the singular point (note that both methods are
expected to give similar results near regular
points~\cite{bender}). In particular, the logarithmic divergence in
the second derivative of the adiabatic Bardeen potential at the NEC
transition is reobtained for any model in which this transition
occurs. We then rephrase, in more general terms, the result that there
is a class of models with NEC transition in which the full Bardeen
potential and all its derivatives are completely regular at all times,
but whose second derivative of the adiabatic perturbation at the NEC
transition is divergent. Finally, we complete this section by
examining the properties of the curvature
perturbation~\cite{bardeen,curvature,lyth} on uniform density
hypersurfaces associated to adiabatic perturbations.

\subsection{The case for a bounded full Bardeen potential: Fluid by Fluid
decomposition}

We now prove that if the matter content of the model is described by
an arbitrary number $N$ of non interacting fluids with constant
equations of state, then the Bardeen potential and all its derivatives
are regular at all times. Using the linearity of Eqs.~(\ref{EE11})
and (\ref{EE12}), we can decompose the total gravitational
perturbation $\Phi$ as the sum $\Phi = \sum_{i}\Phi_{i}$, and
construct the $N$ sets of decoupled equations
\begin{eqnarray}
\nabla^2 \Phi_{i} - 3 \Hu \Phi_{i}' - 3 \left(\Hu^2 -\Ka \right)
\Phi_{i} &=& 4 \pi G a^2 \delta \epsilon_{i},\label{flu1} \\ \Phi_{i}''
+ 3 \Hu \Phi_{i}' + \left( 2 \Hu'+\Hu^2-\Ka\right) \Phi_{i} &=& 4 \pi
G a^2 \delta p_{i},\label{flu2}
\end{eqnarray}
for each value of $i=1,\cdots,N$, \ie for each fluid. Note at this
point that the decomposition of $\Phi$ as the sum of functions
$\Phi_{i}$ is merely a mathematical tool, without any particular
physical meaning, to prove the regularity of the the Bardeen
potential. The functions $\Phi_{i}$ prove convenient as they encode
all the information necessary to describe the dynamical system as a
whole.

Substituting $\delta p_{i} = \omega_{i}\delta \epsilon_{i}$ in
Eq.~(\ref{flu2}), and inserting it into Eq.~(\ref{flu1}), one then
obtains the following decoupled equations for each $\Phi_{i}$:
\begin{equation}
\Phi_{i}'' + 3\Hu (1+\omega_{i}) \Phi_{i}' - \omega_i \nabla^2
\Phi_{i} + \left[ 2\Hu' + \left(\Hu^2 -\Ka\right)
\left(1+3\omega_{i}\right)\right]\Phi_{i} = 0,\label{apg}
\end{equation}
Hence, $2N$ initial conditions are necessary to obtain the full
Bardeen potential; this can also be seen by counting the number of
degrees of freedom and constraint equations~\cite{mfb}.

Models of this type, as proven above, are symmetric around the bounce.
Since $\Hu$ and $\Hu '$ are regular everywhere, and as the
$\omega_{i}$ are constants, then the coefficients appearing in
Eqs.~(\ref{apg}) are completely regular at all times, in particular
around the bounce and around a possible point of NEC transition. Then,
by Fuchs property~\cite{bender}, ${\Phi_{i}}$, and all their
derivatives are also regular at all points.  Consequently, the same
applies true for the full Bardeen potential $\Phi = \sum_{i}\Phi_{i}$.

Note that Eqs.~(\ref{flu1}) and (\ref{flu2}) can be cast in the form
of Eqs.~(\ref{apg}) if and only if the fluid perturbations satisfy
constant equations of state, $\delta p_{i} = \omega_{i} \delta
\epsilon_{i}$, with $\omega_{i}=$const., \ie if the fluids have
vanishing intrinsic entropy perturbations; this most crucial
assumption is not always emphasized.  In the case where the equations
of state contain $\omega_{i}$ which are not constants,
Eqs.~(\ref{apg}) cannot be obtained in this way, the relationship
between $\delta p_{i}$ and $\delta \epsilon_{i}$ may contain divergent
coefficients, as we will see below, and the proof of the regularity of
the Bardeen potential, if possible at all, would be, at least, much
more involved. Note also that if the parameters $\omega_i$ are allowed
to vary, one cannot guarantee that $a(\eta)$ is even, and that the
expansion $a(\eta)=a_0 + b\eta ^2 + e\eta ^4 + \cdots$ is valid near
the bounce.

\subsection{Unboundedness of adiabatic perturbations}

We now investigate the more usual, although alternative in the
bouncing context, description of the perturbations based on an
expansion into adiabatic and entropy modes. Such an expansion works
perfectly well in all other known situation encountered in
cosmology~\cite{mfb}, and it is therefore of interest to understand
whether it can be rendered meaningful during a bouncing phase. 

Let us assume the condition, known to be valid for most fluids, that
the pressure depends on two parameters only, namely the energy density
$\epsilon$, and the entropy $S$. This allows the expansion
\begin{equation}
\delta p = \cS^2 \delta\epsilon+ \bar\tau \delta S,
\label{defmodes}
\end{equation}
where
\begin{equation}
\cS^2 \equiv \left( \frac{\partial p}{\partial \epsilon}\right)_S =
\frac{p'}{\epsilon'}= -\frac{1}{3} \left( 1+
\frac{\beta'}{\Hu \beta}\right)
\label{soundg}
\end{equation}
is the ``sound velocity'' (see Ref.~\cite{nobounce} for a discussion
of this quantity, named $\Upsilon$ in that reference), and $\bar\tau\equiv
(\partial p /\partial S)_\epsilon$. Plugging Eq.~(\ref{EE12}) into
(\ref{EE11}), and making use of the expansion (\ref{defmodes}), one
recovers the usual Bardeen equation
\begin{equation}
\Phi''+ 3\Hu (1+\cS^2) \Phi' + \left[ 2\Hu' + \left(\Hu^2
-\Ka\right) \left(1+3\cS^2\right) - \cS^2 \nabla^2 \right]
\Phi= \frac{3}{2} \lP^2 a^2 \left(\bar\tau \delta S\right).\label{bard}
\end{equation}
{}From now on, we shall Fourier decompose the Bardeen potential and
any other relevant space-dependent quantities on the basis of the
eigenfunctions of the Laplace-Beltrami operator as,
\begin{equation}
\Phi = \sum_k u_k ({\bf x}) \Phi_k(\eta), \hbox{ with } \qquad
(\nabla^2 + k^2) u_k =0,
\end{equation}
where the eigenvalues $k$ depend on the spatial curvature
$\Ka$~\cite{mfb}. In what follows, we shall assume that such an
expansion has been done for all the quantities involved, which will
then be subsequently identified with their Fourier modes, \ie we shall
not write the index $k$, which will be implicit, and make the
replacement $\nabla^2 \to -k^2$ throughout.

The evolution equation for the adiabatic modes is given by
Eq.~(\ref{bard}) in which one sets $\delta S=0$ and
$\Phi\to\Phi_\mathrm{ad}$, namely
\begin{equation}
\Phi_\mathrm{ad}'' + 3\Hu (1+\cS^2 ) \Phi_\mathrm{ad}'
+ \big[ \cS^2 k^2 + 2\Hu' + \left(\Hu^2-\Ka\right) \left(
1+3\cS^2\right)\big] \Phi_\mathrm{ad} =0.\label{usual}
\end{equation}
The presence of $\cS^2$, coming from the expansion (\ref{defmodes}),
in both Eq.~(\ref{bard}) and its equivalent for the purely adiabatic
modes Eq.~(\ref{usual}), together with the fact, as we will see below,
that this quantity may diverge both at the bounce and at the NEC
transition point, indicate some possible bad behavior of $\Phi$ around
these points. As we will see, this is true for Eq.~(\ref{usual}) but
not for Eq.~(\ref{bard}).

It is well known that Eq.~(\ref{usual}) can be solved by means of
the following usual change of variable~\cite{mfb}
\begin{equation} \Phi_\mathrm{ad} = \frac{3 \lP ^2\Hu u}{2 a^2 \theta},
\qquad \qquad \theta \equiv \frac{\Hu}{a}
\sqrt{\frac{3}{2\beta}},\label{Phimu}
\end{equation}
which transforms the original equation into the parametric oscillator
equation
\begin{equation} u'' + \left( \cS^2 k^2 -
\frac{\theta''}{\theta}\right) u=0. \label{u}
\end{equation}
The general solution of this last equation can be constructed
iteratively in the regime for which $\cS^2 k^2\ll
V_u\equiv\theta''/\theta$ to yield
\begin{eqnarray}
\frac{u}{\theta} &=& B_1 \biggl[ 1-k^2\int ^{\eta } \frac{\mathrm{d}\tau
}{\theta ^2} \int ^{\tau }\mathrm{d}\sigma(\cS\theta)^2 +
k^4\int ^{\eta } \frac{\mathrm{d}\tau }{\theta ^2} \int ^{\tau }{\rm
d}\sigma(\cS\theta)^2 \int ^{\sigma} \frac{\mathrm{d}\rho}{\theta ^2}
\int ^{\rho}\mathrm{d}\varsigma(\cS\theta)^2 \biggr]\nonumber\\
&&+B_2\int^{\eta }\frac{\mathrm{d}\tau }{\theta ^2} \biggl[1-k^2\int
^{\tau }\mathrm{d}\sigma(\cS\theta)^2 \int ^{\sigma}\frac{{\rm
d}\rho}{\theta ^2} + k^4 \int ^{\eta } {\rm
d}\sigma(\cS\theta)^2 \int ^{\sigma}\frac{\mathrm{d}\rho}{\theta ^2}\int
^{\rho}\mathrm{d}\varsigma (\cS\theta)^2 \int ^{\varsigma }\frac{{\rm
d}\varrho}{\theta ^2} \biggr]\; \nonumber \\ &&+\cdots ,
\label{musol}
\end{eqnarray}
where $B_1$ and $B_2$ are constants, although usually depending on the
scale $k$ (they are in principle calculated through a matching with
the region where $\cS^2 k^2\gg \theta''/\theta$~\cite{mfb}), and the
remaining terms represented by the dots are of order ${\cal O}(k^6)$
compared with those indicated.

We will now reobtain the divergences we got in Ref.~\cite{nobounce},
this time using expansion (\ref{musol}).

\subsubsection{The bounce}

For the bounce itself, we use expansion (\ref{abounce}) without the
term $f\eta ^{2n+3}$, which is unnecessary here. All the following
behaviors are written up to first order.

The possible cases are:

{\it (1)} $n>1$ and $\Ka \neq 0$.  In this case, we find the following
behaviors (for details, see Ref.~\cite{nobounce}),
\begin{eqnarray}
\cs &=& \frac{2(2n-1)(n-1)}{3\Ka \eta ^2}, \label{cs21} \\ \Hu &=&
\frac{2nb}{a_0}\eta ^{2n-1}, \label{H1} \\ \beta &=&\Ka, \label{beta1}
\\ z&\propto& \frac{1}{\eta ^{2n-2}}\label{z1},
\\ \theta&\propto& \eta ^{2n-1}\label{t1},
\\ V_u&\propto& \frac{1}{\eta ^{2}}\label{v1}\; .
\end{eqnarray}
One can see that $\cs$ diverges with the same power as $V_u$.  Hence,
as long as $k^2 \ll 1$, the expansion (\ref{musol}) can be applied as
close to the bounce as we want. Inserting Eqs.  (\ref{cs21}),
(\ref{t1}) and (\ref{H1}) into (\ref{musol}) and (\ref{Phimu}), one
can easily find that
\begin{equation}
\Phi_\mathrm{ad}\propto \eta ^{2-2n}\label{phi11}\; ,
\end{equation}
which diverges at the bounce, as stated in Ref.~\cite{nobounce}. This
divergence already appears in the term of order $k^0$, and it is
independent of $k$.

{\it (2)} $n>1$ and $\Ka = 0$.  For the special case of a flat
background, the various quantities needed to describe perturbations
are modified as
\begin{eqnarray}
\cs &=& - \frac{a_0(n-1)}{3bn \eta ^{2n}},\label{cs22} \\ \Hu &=&
\frac{2nb}{a_0}\eta ^{2n-1},\label{H2} \\ \beta &=&
-\frac{b}{a_0}2n(2n-1)\eta ^{2n-2},\label{beta2} \\ z &\propto& {\rm
const.}\label{z2}
\\ \theta&\propto& \eta ^{n}\label{t2},
\\ V_u&\propto& \frac{1}{\eta ^{2}}\label{v2}\; .
\end{eqnarray}
Here, $\cs$ diverges faster than $V_u$.  Hence, expansion
(\ref{musol}) cannot be applied as we approach the bounce. In this
situation, one is obliged to approximate the equation and find its
solution. However, in this case, a divergence already in the adiabatic
perturbation itself is found in Ref.~\cite{nobounce},
\begin{equation}
\Phi_\mathrm{ad}\propto \eta^{(3n-2)/2} \hbox{e}^{|\alpha|},\label{phi12}\;
\end{equation}
where
\begin{equation}
\alpha \equiv k\sqrt{a_0 / [3 n b (n-1)]}\eta^{1-n}\; . \nonumber
\end{equation}
Note there is no divergence for $k=0$.

{\it (3)} $n=1$, $d\neq 0$, $\forall\Ka$.  This is the case where the
second derivative of $a(\eta)$ is non vanishing and $a(\eta)$ is not
even. The relevant quantities are
\begin{eqnarray}
\cs &=& -\frac{a_0d}{b (2b-\Ka a_0)\eta},\label{cs23} \\ \Hu &=&
\frac{2b}{a_0}\eta,\label{H3} \\ \beta &=&\Ka -
\frac{2b}{a_0},\label{beta3} \\ z &\propto& \frac{1}{\sqrt{\eta}}
\label{z3}\\ \theta&\propto& \eta \label{t3},
\\ V_u&\propto& \frac{1}{\eta}\label{v3}\; .
\end{eqnarray}
As in the first case, $\cs$ diverges with the same power as $V_u$.
Again, as long as $k^2 \ll 1$, the expansion (\ref{musol}) can be
applied as close to the bounce as we want. Inserting Eqs.
(\ref{cs21}), (\ref{t1}) and (\ref{H1}) into (\ref{musol}) and
(\ref{Phimu}) one can easily find that the adiabatic perturbation is
finite but its first derivative diverges logarithmically, $\Phi'_{\rm
ad}\propto B_2 k^2\ln(\eta)$, and its second derivative diverges as
$\Phi''_\mathrm{ad}\propto B_2 k^2/\eta$, exactly as stated in
Ref.~\cite{nobounce}. There is no divergence for $k=0$.

{\it (4)} $n=1$, $\forall \Ka$, and $d=0$.

This is the case we will examine in detail in this paper.  The
relevant quantities are
\begin{eqnarray}
\cs &=& \frac{8b^2+(\Ka b - 12 e)a_0}{3b (2b-\Ka a_0)},\label{cs24} \\
\Hu &=& \frac{2b}{a_0}\eta,\label{H4} \\ \beta &=&\Ka -
\frac{2b}{a_0},\label{beta4} \\ z &\propto& \frac{1}{\eta}
\label{z4}\\ \theta&\propto& \eta \label{t4},\\ V_u &=& \mathrm{const.}
\label{v4}\;
\end{eqnarray}
In this case, neither $\cs$ nor $V_u$ diverge at the bounce. Hence,
Eq.~(\ref{usual}) is regular around the bounce, and so are all its
solutions\footnote{Note that, in all these discussions, we are not
examining situations where the constants appearing in the relevant
quantities exhibited above cancel out exactly, as this requires some
fine tuning.}.

\subsubsection{The NEC transition}

In what follows, we concentrate on the point where $\beta=0$, so that
we shift the origin of time~: for the rest of this section, $\eta=0$
when $\beta=0$, and we denote by an index $0$ quantities evaluated at
this point.

We now assume that the scale factor around $\eta=0$ is, again,
differentiable at least up to third order, so that the following
expansion
\begin{equation} a(\eta) = a_0 \left[ 1+\Hu_0\eta +\frac{1}{2}
\left( 2\Hu_0^2 +\Ka\right) \eta^2 + \frac{1}{3!} a_3 \eta^3 + \cdots
\right],
\label{a0}\end{equation}
holds.\footnote{Apart from the fine tuned case having $2b=a_0$, the
situation for which the point of NEC transition coincides with the
bounce [\ie no linear term in the expansion (\ref{a0})] is nothing but
the {\it case (2)} of the previous subsection, which is thus already
treated.}  In this relation, $a_3 \equiv a'''(0) /a_0$, and use has
been made of $a''(0)/a_0 = 2\Hu_0^2+\Ka$, which is a simple
rewriting of $\beta = 0$.

Using the expansion~(\ref{a0}), we find that
\begin{equation}
\Hu = \Hu_0+\eta (\Hu_0^2+\Ka) +\frac{1}{2}
\left[ a_3 - \Hu_0\left( 4 \Hu_0^2+3 \Ka\right)\right] \eta^2
+{\cal O}(\eta^3),
\end{equation}
leading to
\begin{eqnarray}
\beta &\simeq& \left[ \Hu_0 \left( 6\Hu_0^2+5 \Ka\right)
-a_3\right] \eta + {\cal O} \left( \eta^2 \right),
\end{eqnarray}
while the sound velocity takes the form
\begin{eqnarray}
\cs &= & - \frac{1}{3 \Hu_0 \eta} + {\cal O} (\eta^0),
\end{eqnarray}
which manifestly diverges at the NEC violating point.

We will now use Eqs.~(\ref{Phimu}) and (\ref{musol}) to evaluate the
divergences in the adiabatic perturbation. The relevant quantities in
these equations are, to leading order
\begin{eqnarray}
\label{te-2} \theta^{-2} &=& \frac{2a_0^2}{3\Hu_0}\left[
\Hu_0\left(6\Hu_0^2+5\Ka\right) - a_3\right] \eta + {\cal O}
(\eta^2),
\end{eqnarray}
whose behavior, combined with the divergence in $\cs$, yields
\begin{equation}
\label{c2te} \cs\theta^2 \simeq
\frac{\Hu_0\eta^{-2}+\Ka\eta^{-1}}{2 a_0^2 \left[
  a_3 -\Hu_0 \left(6\Hu_0^2+5\Ka\right) \right]}
  \equiv b_1\eta^{-2}+b_2\eta^{-1}+{\cal O} (\eta^0)\; ,
\end{equation}
and finally
\begin{equation}
\label{ha2} \frac{\Hu}{a^{2}} =
\frac{\Hu_0}{a_0^2}+\frac{(\Ka-\Hu_0^2)}{a_0^2}\eta +\frac{[a_3- 3
\Hu_0\left(2 H_0^2+3\Ka\right)]}{2a_0^2}\eta^2,
\end{equation}
plus terms of order ${\cal O} (\eta^3)$.

Inserting these expressions into Eqs.~(\ref{Phimu}) and (\ref{musol}),
one obtains, in the terms with coefficient $B_1$, the quantities
$k^2b_2c_1\eta^2\ln(\eta)/2$ and $-k^4 b_1^2c_1^2\eta^2\ln(\eta)/2$,
whose second derivative diverges as $\ln(\eta)$. This is exactly the
type of divergence obtained in Ref.~\cite{nobounce} by another method.
Here, however, the $k$-dependence of the divergences is obtained more
precisely. There is no divergence for $k=0$.  Note that if $\Ka=0$,
the coefficient $b_2$ vanishes, so that the divergence appears only at
order $k^4$. We will return to this point in the last section.

The divergence in the adiabatic perturbation presented above, which is
present in any model with NEC transition points (including the class
of the precedent subsection), suggests that adiabatic perturbations
cannot be defined in such models. We will turn to this point in
details in Sec.~\ref{sec:pertgen2} by concentrating on the simplest
situation involving only two fluids. In the meantime, let us end up
the setting of the general formalism for $N$ fluids by looking at the
curvature perturbation.

\subsection{Curvature perturbation in the adiabatic case}\label{sec:zeta}

Another relevant function that can be useful for calculating the
primordial spectrum of cosmological perturbation was introduced in
Ref.~\cite{curvature}. It is the curvature perturbation $\zeta$ on
uniform density hypersurfaces (or its generalization for non-flat
background $\zeta_{_\mathrm{BST}}$~\cite{zBST})
\begin{equation}
\zeta \equiv
\frac{2}{3}\biggl(\frac{\Hu^{-1}\Phi'+\Phi}{1+\omega}\biggr)+\Phi =
\frac{\left(\Hu\Phi\right)'}{\beta} + 2 \Phi,\label{zetadef}
\end{equation}
where $\omega$ is the total equation of state which can be evaluated
by means of Eqs.~(\ref{EE01}) and (\ref{EE02}) through
\begin{equation}
1+\omega = \frac{\epsilon+p}{\epsilon} = \frac{2}{3}
\biggr(\frac{\beta}{\Hu^2 +\Ka}\biggl) \ ,
\end{equation}
and the second equality of Eq.~(\ref{zetadef}) stems from this
relation together with the explicit assumption $\Ka=0$.

This variable is useful in particular when it comes to describing
ordinary transitions such as the radiation to matter domination, or
the reheating at the end of inflation: on large (super-Hubble, often
misleadingly called superhorizon~\cite{jmpp}) scales, $\zeta$ is
approximately constant. Indeed, in this case, it is found
that~\cite{lyth}
\begin{equation}
\zeta' \simeq -\frac{\Hu}{\epsilon + p} \delta p_\mathrm{nad},
\label{zetaprime}
\end{equation}
where $\delta p_\mathrm{nad} = \delta p -\cs \delta \epsilon$ is the
nonadiabatic part of the pressure perturbation, so that $\zeta$ is
expected to be conserved for adiabatic perturbations.  In this latter
case, to which we restrict our attention in this section, it therefore
suffices to evaluate it before the transition to obtain its value
after the transition, without prior knowledge of the detailed
structure of the transition itself. It is immediately clear however
from Eq.~(\ref{zetaprime}) that this will no longer hold whenever a
nonnegligible amount of entropy perturbation is present, or if the NEC
violation occurs at some stage. As we have seen above, both these
conditions generically take place in a bouncing scenario, so that care
must be taken in examining this particular case. In a fashion similar
to Eq.~(\ref{Phimu}), one can define
\begin{equation}
\zeta_\mathrm{ad}=-\sqrt{\frac{3}{2}}\lP \frac{v}{z},\label{zetav}
\end{equation}
where $v$ gives yet another way of obtaining the gravitational
potential $\Phi$ through~\cite{mfb}
\begin{equation}
\Phi_\mathrm{ad} =\sqrt{\frac{3}{2}} \frac{\lP \beta^{1/2} z}{a
\cS k^2} \left(\frac{v}{z}\right)', \label{phizv}
\end{equation}
with
\begin{equation}
z\equiv \frac{a \beta^{1/2}}{\Hu \cS} =
\sqrt{\frac{3}{2}}\frac{1}{\cS\theta}, \label{zdef}
\end{equation}
where, again, care must be taken when $\beta$ and $\cS^2$ change sign,
which occurs at the NEC transition. Note that Eq.~(\ref{zetav}) is
only valid provided one considers adiabatic perturbations; if entropy
perturbations were present at a nonnegligible level, then
Eq.~(\ref{zetav}) would have to be modified by inclusion of an extra
term, proportional to the entropy perturbation in question (and to
$z$), in order to define the full $\zeta$.

It is worth noting at this point that the variable $v$ draws its
importance in the theory of cosmological perturbations from the fact
that, in the case of a single fluid (or a scalar field) dominating the
universe, it allows to write the total (gravitational and fluid)
action as that of a simple scalar field with varying mass, which can
then be easily quantized~\cite{mfb}.  It is from this variable, by
assuming a Bunch-Davies~\cite{BD} vacuum state for its relevant modes,
that one sets initial conditions for the cosmological perturbation.

The effective action derivable for $v$ permits to cast its equation of
motion in the same form as Eq.~(\ref{u}), namely
\begin{equation} v'' + \left( \cS^2 k^2 -
\frac{z''}{z}\right) v=0, \label{v}
\end{equation}
where now the effective potential reads
\begin{equation}
V_v\equiv \frac{z''}{z},\label{Vv}
\end{equation}
and the RHS of Eq.~(\ref{v}), although vanishing in the adiabatic
case to which we restrict our attention here, contains in
principle a source term proportional to the entropy perturbation.
The perturbation $v$, being commonly used in the literature as a
quantum scalar field, may be argued to possess some amount of
physical significance, although it is not directly observable.
Note also that $v$ cannot account for all the degrees of freedom
if more than one fluid are acting on comparable levels; in this
latter case, the variable $v$ merely encodes the information on
the adiabatic part of the full Bardeen potential.

The solution can be similarly expanded as in Eq.~(\ref{musol}) for the
variable $u$, provided the replacements $u\to v$ and $\theta\to z$ are
done. Using Eq.~(\ref{phizv}), this gives the curvature perturbation
directly as
\begin{equation}
\zeta_\mathrm{ad}\propto\frac{v}{z} \simeq C_1 \left[ 1-k^2\int
^{\eta } \dd\tau (\cS\theta)^2 \int ^{\tau }
\frac{\dd\sigma}{\theta ^2}
\right]
+C_2\int ^{\eta } \dd\tau (\cS\theta)^2 \left[1-k^2\int ^{\tau}
\frac{\dd\sigma }{\theta ^2} \int ^{\sigma}\dd\rho(\cS\theta)^2
\right],\label{vsol}
\end{equation}
where $C_1$ and $C_2$ are again constants depending on scale and we
have dropped higher order terms.

In principle, using either $u$ or $v$ to propagate adiabatic
perturbations through a given period should lead to the same
gravitational adiabatic potential $\Phi_\mathrm{ad}$, and hence to the
same primordial spectrum, \ie the same physical predictions. In
practice however, because of the existence of poles in the effective
potentials and ``sound velocity'', there are instances leading to
discrepancies (see Refs.~\cite{jmpp,cdc} for a more thorough
discussion).

The results obtained in this section suggest that the very definition
of adiabatic modes may be unattainable in such bouncing models. In
fact, as explained in Sec.~\ref{sec:back}, one needs at least two
fluids in order to construct a realistic bouncing model. This, in
turn, demands that the perturbations cannot be purely adiabatic: even
if one considers fluids with no intrinsic entropy perturbation, the
relative entropy contribution must be present. The question is if the
entropy fluctuations, although necessarily present, can be considered
to be irrelevant all along in order for the adiabatic perturbations to
be sufficient to accurately describe the evolution of the overall
perturbations, as usually assumed to be the case on large scales. In
order to answer this question, we examine in details in the following
section (Sec.~\ref{sec:pertgen2}) a prototypical example of a two non
interacting fluids bouncing model, both fluids having barotropic
equations of state with constant ratio.

\section{Bouncing with two fluids}\label{sec:pertgen2}

In what follows, we establish the differential equations governing the
entropy perturbations and the Bardeen potential dynamical evolution,
and show that entropy fluctuations cannot be neglected at the NEC
transition point, and hence that purely adiabatic perturbations have
no meaning there. To conclude this section, \ie in \ref{sec:pertspec},
we restrict ourselves to the particular case of a flat universe filled
with a negative energy stiff matter and radiation, or, in other words,
the situation equivalent to the case presented in
Ref.~\cite{ppnpn2}. There we show explicitly that, although presenting
the abovementioned divergences in the adiabatic fluctuations, the
equations governing the Bardeen potential and entropy fluctuations,
when taken together without any adiabatic condition, yield a perfectly
regular fourth order equation for the full Bardeen potential, which is
equal to the one obtained in Ref.~\cite{ppnpn2}, whose solutions and
derivatives must be regular at any time, including at the NEC
transition. This reinforces the idea that there is no problem in
defining entropy fluctuations in such bouncing models, but that we
cannot neglect them around the NEC transition point.  We also study,
in this particular example, the behavior of curvature perturbations.

\subsection{Adiabatic and entropy modes}

We now specialize to the $N=2$ case for which many calculations,
in particular the evolution of entropy perturbations, can be done
explicitly. We now have $\delta \epsilon = \delta \eun +\delta
\edeu$ and $\delta p = \delta p_1 + \delta p_2$, from which we
can derive the constraint~\cite{mfb}
\begin{equation}
(a \Phi)'_{;i} = 4 \pi G a^2 \left[ \eun (1+\wun ) \delta u_{(1) i} +
\edeu (1+\wdeu ) \delta u_{(2) i}\right] ,\label{EEcont}
\end{equation}
which will be useful later. The sound velocity now takes the simple
form
\begin{equation}
\cS^2 = \frac{\wun (1+\wun)\eun + \wdeu (1+\wdeu) \edeu}{
(1+\wun) \eun + (1+\wdeu) \edeu},
\label{sound}
\end{equation}
which diverges at the points of NEC violation $\epsilon + p =0$.  With
these explicit relations, one can calculate the entropy contribution
to the pressure fluctuation. This is
\begin{equation}
\bar\tau\delta S =(\wun-\wdeu)
\frac{(1+\wun)(1+\wdeu)\epsilon_1\epsilon_2}{(1+\wun)\epsilon_1 +
(1+\wdeu)\epsilon_2} \times\times \left[ \frac{\delta\epsilon_1}{
(1+\wun)\epsilon_1} - \frac{\delta\epsilon_2}{(1+\wdeu)\epsilon_2}\right],
\label{deltaS}
\end{equation}
for which we now seek a time evolution equation.

The perturbed fluid tangent vector, for each fluid, has components
given by $\delta u_0 = - a^2 \delta u^0 = a \Phi$ and $\delta u_{(1,2)
i} = - a^2 \gamma _{ij} \delta u_{(1,2)}^j = -a \partial_i v_{(1,2)}$,
thus defining the potentials $v_1$ and $v_2$.  Expanding the
energy-momentum conservation $\nabla_\mu T^{\mu\nu}_i=0$ for both
fluid to first order yields
\begin{equation}
\delta\epsilon_i'+3\Hu (\delta\epsilon_i +\delta p_i) + (\epsilon _i
+p_i) (\nabla^2 v_i -3\Phi') =0,
\label{T0}
\end{equation}
for the time component, and
\begin{equation}
[(\epsilon_i+p_i) \partial_j v_i]'+4\Hu (\epsilon_i +p_i) \partial_j
v_i + \partial_j \delta p_i + (\epsilon_i+p_i)\partial_j \Phi =0,
\label{Tk}
\end{equation}
for a spatial component. Defining the density contrasts through
\begin{equation}
\delta_i\equiv \frac{\delta\epsilon_i}{\epsilon_i},
\end{equation}
and using the background energy momentum conservation, which implies
$\delta_i' = [\delta\epsilon_i'+3\Hu
(1+\omega_i)\delta\epsilon_i]/\epsilon_i$, one transforms
Eq.~(\ref{T0}) into
\begin{equation}
\delta_i' = (1+\omega_i) (3\Phi'-\nabla^2 v_i).
\label{deltai}
\end{equation}

Let us emphasize at this stage the well-known fact that the
adiabaticity condition
\begin{equation}
\delta S = 0 \Longleftrightarrow s\equiv
\frac{1}{\wun -\wdeu} \left(\frac{\delta_1}{1+\wun} -
\frac{\delta_2}{1+\wdeu}\right) =0
\label{adiab}
\end{equation}
is conserved in time in the long wavelength limit $k\to 0$ since
$(\wun - \wdeu) s'= k^2 (v_1 - v_2)$. This is one of the reasons for
considering adiabatic modes, the other being that the observed
spectrum of primordial fluctuations can be reconstructed, \eg from
CMBFAST~\cite{CMBFAST}, with initial conditions deep in the radiation
era satisfying $s=0$, whereas isocurvature modes lead to significant
disagreement~\cite{isocurv} with the data~\cite{dataCMB}, although
some mixture is still acceptable~\cite{mix} (the situation is similar
to that of having a small component of the perturbations in the form
of topological defects~\cite{bprs}).

Projecting Eq.~(\ref{Tk}) along the vector $k_j$ leads to the
following dynamical equation for the velocity potentials $v_i$:
\begin{equation}
v_i'+\Hu (1-3\omega_i) v_i + \omega_i \frac{\delta_i}{1+\omega_i} +
\Phi = 0.\label{vi}
\end{equation}
With the definition (\ref{adiab}) and Eqs.~(\ref{deltai}) and
(\ref{vi}), upon using the constraint equation written as

\begin{equation}
(3\Ka-k^2)\Phi = \frac{3}{2} \lP^2 a^2 \sum_i \left\{
\epsilon_i \delta_i -3 \Hu \left[ \epsilon_i (1+\omega_i) v_i
\right]\right\},
\end{equation}
which is nothing but a rewriting in a convenient way of
Eq.~(\ref{EE11}) using Eq.~(\ref{EEcont}) and the background Einstein
equation~(\ref{EE02}), the dynamical equation for the variable $s$
follows directly. This is:
\begin{equation}
s''+\Hu (1-3\cZ^2)s' + k^2 \cZ^2 s = \frac{k^2}{\beta} (k^2 - 3\Ka)
\Phi,
\label{eqS}
\end{equation}
where the definition~\cite{cZ}
\begin{equation}
\cZ^2 \equiv \frac{\wdeu \eun (1+ \wun) + \wun \edeu (1+
\wdeu)}{(1+\wun) \eun + (1+\wdeu) \edeu} \label{cZ2}
\end{equation}
has been used. It should be noted that this function, just as the
``sound velocity'' $\cS^2$, is also singular at the NEC violating
point, for which $(\epsilon + p)\to 0$.

Eq.~(\ref{eqS}) is the main result of this section. Note that its
source term diverges at the NEC transition point. It means that we
cannot neglect $s$ at this point for any small but finite $k$.
Hence, adiabatic perturbations cannot be defined there. Note also
that the divergent source is of order $k^2$ if $\Ka\neq 0$, and
$k^4$ if $\Ka = 0$, exactly the orders in which the divergences
in the adiabatic potential appear as calculated in
section~\ref{sec:genpert}.  Finally, we want to emphasize from
Eq.~(\ref{eqS}) that the adiabatic case is, in most of the usual
situations not involving NEC violation or bounces, \ie in
standard inflationary models, the only one that is tractable self
consistently, in particular in the flat $\Ka=0$ situation. This is
because $s\simeq 0$ solves Eq.~(\ref{eqS}) at leading order in
$k^2$. The symmetric situation, with ``purely entropic modes'',
having $\Phi=0$, is not self-consistent because $s$ [or $\delta
S$ in Eq.~(\ref{bard})] then sources the gravitational potential
at the same order.

Eqs.~(\ref{bard}) and (\ref{eqS}) form a closed system for the
perturbation variables $s$ and $\Phi$ depending only on the background
functions. In fact, one can express the entropy and sound
``velocities'' $\cZ$ and $\cS$ simply as
\begin{equation}
\cZ^2 = \frac{3}{2} \left( 1+\wun+\wdeu+\wun \wdeu \right)
\frac{\Hu^2+\Ka}{\beta} - 1,\label{cZ2aeta}
\end{equation}
and
\begin{equation}
\cS^2 = -\frac{1}{3} \left( 1+\frac{\beta'}{\Hu\beta} \right),
\label{cs2aeta}
\end{equation}
whereas the left hand side of Eq.~(\ref{bard}) can be given the form
\begin{equation}
\frac{3}{2} \lP^2 a^2 (\bar\tau \delta S) = -\frac{\beta
\left(\cS^2\right)'}{3\Hu} s.
\label{source}
\end{equation}
All these relations permit to gather the overall system into a much
simplified form involving only the scale factor $a(\eta)$, system
which is therefore particularly well suited for a more specific
investigation. Let us accordingly now apply these relations to the
particular case of Ref.~\cite{ppnpn2}, which is, as mentioned above,
equivalent to a two fluids model consisting of a negative energy stiff
matter and radiation. We want to verify whether, without any adiabatic
condition, Eqs~(\ref{bard}) and (\ref{eqS}) are consistent with the
general result that the full Bardeen potential and all its derivatives
are bounded at all times.

\subsection{A worked-out example}\label{sec:pertspec}

We now consider the particular case of a bounce occurring for a flat
($\Ka=0$) universe filled with radiation ($\wun =1/3$) and some
negative energy stiff matter having $p_\mathrm{s} = \epsilon_\mathrm{s} <0$
(\ie, $\wdeu =1$). Energy conservation is valid separately for both
fluids, yielding $\epsilon_{\mathrm{r}} = c_{\mathrm{r}}/a^4$ and
$\epsilon_{\mathrm{s}} = - c{_\mathrm{s}}/a^6$, with $c{_\mathrm{r}}$ and
$c_{\mathrm{s}}$ two positive constants. Note that this stiff matter can
also be modeled by a negative kinetic energy free massless scalar
field, whose background dynamics and perturbations were studied in
Ref.~\cite{ppnpn2}.

The background FLRW metric has a scale factor that takes the
form~\cite{ppnpn2}
\begin{equation}
a=a_0 \sqrt{1+\left(\frac{\eta}{\eta_0}\right)^2},
\label{scale}
\end{equation}
where $a_0^2 = c_\mathrm{s}/c_\mathrm{r}$ and $\eta_0^2 =
c_\mathrm{s}/(c_{\rm r}^2 \lP^2)$. Note that this form satisfies the
symmetry requirement $\eta\to -\eta$ proven in section~\ref{sec:back}.

The relevant quantities to be calculated, namely $\Hu$, $\Hu '$,
$\beta$, $\cZ^2$, and $\cS^2$ read, in this example,
\begin{equation} \Hu (x) \equiv \frac{x}{\eta_0 (1+x^2)},\ \ \ \ \Hu '(x)
\equiv \frac{1-x^2}{\eta_0^2(1+x^2)^2},\label{hu2}\end{equation}
\begin{equation} \cS^2 (x) \equiv \frac{2x^2-7}{3(2x^2-1)},\ \ \ \ \cZ^2(x)
\equiv \frac{2x^2+1}{2x^2-1},\label{csz}\end{equation}
\begin{equation} \beta (x) \equiv \frac{2x^2-1}{\eta_0^2(x^2+1)^2},
\label{beta}\end{equation}
where we have set for further convenience $x\equiv\eta/\eta_0$.

The equivalent of Eqs~(\ref{apg}) now read [recall that for the sake
of simplicity, and to avoid an unnecessary proliferation of indices,
we note simply $\Phir$ and $\Ps$ the $k-$modes of $\Phi$ in the
following equations, \ie \eg $\Phir \equiv \Phir(k,\eta)$]
\begin{equation}
\frac{\dd^2\Phir}{\dd x^2} + \frac{4x}{1+x^2}
\frac{\dd \Phir}{\dd x} + \left[ \frac{\tilde k^2}{3} +
\frac{2}{(1+x^2)^2}\right] {\Phir} = 0,
\label{ap3}
\end{equation}
and
\begin{equation}
\frac{\dd^2\Ps}{\dd x^2} +
\frac{6x}{1+x^2}\frac{\dd \Ps}{\dd x} + \left( \tilde k^2 +
\frac{2}{1+x^2}\right) {\Ps} = 0,
\label{ap4}
\end{equation}
where $\tilde k \equiv k\eta _0$, with $\Phi = \Phir + \Ps$, the
indices r and s corresponding to radiation and stiff matter,
respectively.  As one can see, all coefficients in these equations are
completely regular, even at the point of NEC transition,
$\xnec^2=1/2$.  The full Bardeen potential is therefore regular
everywhere.

Let us now examine the evolution of the perturbations from the more
usual alternative point of view of Eqs.~(\ref{bard}) and (\ref{eqS}).
{}From Eqs.~(\ref{source}), (\ref{hu2}), (\ref{csz}), and
(\ref{beta}), one can turn Eqs.~(\ref{bard}) and (\ref{eqS}) into
\begin{eqnarray}
\frac{\dd^2\Phi}{\dd x^2} + \frac{2x(4x^2-5)}{(1+x^2)(2x^2-1)}
\frac{\dd \Phi}{\dd x} &+& \left[ \frac{2x^2-7}{3(2x^2-1)} \tilde k^2 -
\frac{2}{(1+x^2)(2x^2-1)} \right] \Phi \nonumber \\
&=& -\frac{8}{3}\frac{s}{(1+x^2)(2x^2-1)},\label{bard2}
\end{eqnarray}
and
\begin{equation}
\frac{\dd^2 s}{\dd x^2}-\frac{4x}{2x^2-1} \frac{\dd s}{\dd x} +
\frac{2x^2+1}{2x^2-1}\tilde k^2 s =
\tilde k^4 \frac{(x^2+1)^2}{2x^2-1} \Phi.
\label{eqS27}
\end{equation}

The coefficients of $\Phi$, $\dd \Phi /\dd x$, $s$ and $\dd s /\dd x$
in these equations, as well as their source terms, diverge as
$1/(2x^2-1)$ at the NEC transition points $\xnec$.  This may imply
divergences in the solutions of these equations near the NEC
transition. However, taking into account the entropy modes (which was not
possible at all in the framework of Ref.~\cite{nobounce}) without any
adiabatic assumption allows a more precise analysis
since it permits to obtain the fourth order equation satisfied by {\sl
all} the modes of $\Phi$, \ie not only the adiabatic ones (whose
meaning near NEC transition is in question). The relevant equation,
derived from Eqs.~(\ref{bard2}) and (\ref{eqS27}) is
\begin{equation}
\frac{\dd^4\Phi}{\dd x^4} + \frac{10x}{1+x^2} \frac{\dd^3\Phi}{\dd x^3} +
\left(\frac{4}{3} \tilde k^2 +
\frac{20}{1+x^2}\right)\frac{\dd^2\Phi}{\dd x^2}+ \frac{6x\tilde
k^2}{1+x^2}\frac{\dd \Phi}{\dd x} + 
\frac{1}{3}\left(\tilde k^2+\frac{4}{x^2+1} \right) \tilde k^2 \Phi
= 0. \label{Phix4}
\end{equation}
All the coefficients of the above equation, as could have been
expected from the analysis in terms of Eqs.~(\ref{ap3}) and
(\ref{ap4}), are regular everywhere. Hence, using again Fuchs
property, $\Phi$ and all its derivatives must be regular at all
points, \ie there cannot be any divergence nowhere, no unbounded
growth of the perturbation, in accordance with the alternative
analysis based on Eqs.~(\ref{ap3}) and (\ref{ap4}).  Using
Eq.~(\ref{hu2}), one can show that Eq.~(\ref{Phix4}), as expected, is
the same as the one obtained in Ref.~\cite{ppnpn2} [Eq.~(35) thereof],
where the negative energy stiff matter is described in terms of the
negative energy free massless scalar field, namely
\begin{equation}
\Phi^{^{\mathrm (IV)}}+10\Hu\Phi'''+\left[ \frac{4}{3} k^2 + 20
\left( \Hu'+2\Hu^2 \right) \right]\Phi''+ 6 \Hu k^2
\Phi'+\frac{1}{3}k^2 \left[ k^2 + 4\left( \Hu'+2\Hu^2 
\right) \right]\Phi =0.
\label{Phi4}
\end{equation}

In the same way one can obtain, for the sake of completeness, a
similarly decoupled fourth order equation for the entropy $s$, which
reads
\begin{equation}
s^{^{\mathrm (IV)}}-2\Hu s'''+\left[ \frac{4}{3} k^2 + 2
\left( \Hu'+2\Hu^2 \right) \right]s''
- \frac{2}{3} \Hu k^2 s'+
k^2 \left[ \frac{1}{3} k^2 -2 \left( \Hu'+2\Hu^2 \right) \right]s =0.
\label{s4}
\end{equation}
Hence, as $\Hu'$ and $\Hu$ are regular functions, the entropy
fluctuations are also well behaved all along.

It should be kept in mind that Eqs.~(\ref{Phi4}) and (\ref{s4}) do in
fact represent the same physical system, and the second does not
provide any additional information not already contained in the
first. The number of initial conditions, which is obtained from only
one of them, is four, as one should expect in the particular case of
two fluids [see the discussion below Eq.~(\ref{flu2})].

Let us now investigate the pure adiabatic ($s=0$) perturbations modes.
Their equation is given by Eq.~(\ref{usual}), which we solve by means
of Eqs.~(\ref{Phimu}) and (\ref{musol}). Note that although the term
$\cS^2$ presents a (simple) pole at $x=\xnec$, the approximation
(\ref{musol}) makes perfect sense even around this point since the
effective potential for $u$ reads
\begin{equation}
V_u\equiv \frac{\theta''}{\theta} = \frac{8 x^6 -2x^4+20 x^2+3}{\left[
\eta_0 \left(1+x^2 \right) \left(2 x^2-1\right)\right]^2},
\label{Vu}
\end{equation}
and hence presents two second order poles: the approximation actually
becomes better as one gets closer to the NEC transition points.

The relevant quantities are, to leading order,
\begin{equation}
\label{c2tep}
\cs\theta^2 = \frac{x^2(2x^2-7)}{2a_0^2(1+x^2)(2x^2-1)^2}
\simeq-\frac{1}{8a_0^2}\left[\frac{1}{\left( x-\xnec \right)^2} -
\frac{13}{6} \right],
\end{equation}
the following term being of order ${\cal O} \left( x-\xnec \right)$,
and
\begin{equation}
\label{te-2p}
\theta^{-2} = \frac{2a_0^2}{3x^2}\left( 1+x^2\right)\left(
2x^2-1\right) \simeq 4 a_0^2\left[\sqrt{2}\left( x-\xnec\right)-
\frac{5}{3}\left(x-\xnec\right)^2 \right],
\end{equation}
where now the remaining term behaves as ${\cal O} \left[\left( x-\xnec
\right)^3\right]$. These relations are precisely of the form expected
from Eqs.~(\ref{c2te}) and (\ref{te-2}) for $\Ka =0$, \ie with $b_2 =
0$.

The integrals in Eq.~(\ref{musol}) can be evaluated explicitly, and
read
\begin{equation}
\int^\eta \frac{\dd\tau}{\theta^2} = \frac{2}{3} a_0^2 \eta_0 \left(
\frac{1}{x} + x + \frac{2}{3} x^3\right),
\end{equation}
and
\begin{equation}
\int^\eta \cS^2 \theta^2 \dd\tau = \frac{\eta_0}{2a_0^2} \left(
\frac{x}{2x^2 -1} + \arctan x\right),
\end{equation}
which in turn yield
\begin{equation}
\label{intsp}
\int^\eta \frac{\dd\tau}{\theta^2}\int^\tau \cS^2\theta^2 \dd\sigma =
\frac{\eta_0^2}{18 x} \left\{ x^3 + 2\left[ 2x^4+3 (x^2+1)\right]
\arctan x + 2 x \ln (x^2+1)\right\}, 
\end{equation}
and
\begin{equation}
\int^\eta \cS^2\theta^2 \dd\tau \int^\tau\frac{\dd\sigma}{\theta^2} =
\frac{\eta_0^2}{18 (2x^2-1)} \left[ 10 - x^2 + 2 x^4 + 2(1-2x^2)
 \ln(x^2+1)\right]. 
\end{equation}
The divergence in the second derivative appears in the term
\begin{equation}
\Phi_\mathrm{ad, dom}'' \sim  B_1 k^4 \left[
\cs \int^\eta \frac{\dd\tau}{\theta^2}\int^\tau \cS^2\theta^2 \dd\sigma
+ (\theta ^{-2})'
\int^\eta \cS^2\theta^2 \dd\tau\int^\tau \frac{\dd\sigma}{\theta^2}
\int^{\sigma} \cS^2\theta^2 \dd\rho\right], \label{terms}
\end{equation}
which, after inserting Eqs~(\ref{c2tep}), (\ref{te-2p}), (\ref{intsp})
and (\ref{csz}) expressed as Taylor series around $x-\xnec$
yields
\begin{equation}
\label{divnec} \Phi_\mathrm{ad, dom}''\propto B_1 k^4
\ln\left(x-\xnec\right),
\end{equation}
as expected from the results of section~\ref{sec:genpert}. 

Note that in order to get Eq.~(\ref{divnec}), it is necessary to keep
the subleading part in the second term of Eq.~(\ref{terms}) since the
leading orders of both terms are $\propto \left( x-\xnec \right)^{-1}$
and exactly cancel each others. This result, together with the regular
equations (\ref{Phix4}) and (\ref{s4}) obtained above, show that the
Bardeen potential and the entropy perturbations given as solutions of
Eqs.~(\ref{bard2}) and (\ref{eqS27}) make sense at the NEC transition,
but that the adiabatic modes in themselves do not. However, since the
divergence itself arises at the fourth order in $k$, and in the second
derivative of the adiabatic part of the Bardeen potential, it can be
argued that, in the flat case $\Ka=0$ at least, this adiabatic part of
the Bardeen potential and its first derivative could be used in a
consistent way to produce matching conditions in a bouncing scenario.

To be complete, let us now compute the curvature perturbation as
discussed in Sec.~\ref{sec:zeta}. We first evaluate the background
function $z$ in the background (\ref{scale}). This gives
\begin{equation}
z=\frac{a_0}{x} \left( 2x^2-1 \right) \left[ \frac{3 \left(
1+x^2 \right) }{2x^2-7}\right]^{1/2},
\label{z}
\end{equation}
which in turns yields the following complicated form for the potential $V_v$
\begin{equation}
V_v=\frac{14 x^6 +108 x^4 +273 x^2 + 98}{\left[ \eta_0 x \left( x^2+1
\right) \left(2x^2-7\right) \right]^2},
\end{equation}
and thus exhibits two second-order poles whose origin can be traced
back to the time at which the ``sound velocity'' $\cS$ vanishes, and
another pole at the bounce itself. As is apparent from the solution
(\ref{vsol}), however, the points at which $\cS\to 0$ do not generate
any divergences, and are in fact perfectly regular as far as the
solution is concerned; therefore, we shall not consider them further.

The solution given by Eq.~(\ref{vsol}) involves essentially the same
integrals as those in Eq.~(\ref{musol}), except for the last one (the
triple integral), which we could not perform explicitly. It turns out,
however, that, contrary to the Bardeen potential, and because of the
reverse order in which the integrands appear in the integrals of
Eq.~(\ref{vsol}) with respect to Eq.~(\ref{musol}), the curvature
perturbation diverges at the NEC transition, and is thus not a proper
quantity to propagate through this point. Since $\Phi_\mathrm{ad}$, the
expansion rate $\Hu$, and their first derivatives are all well-behaved
at the NEC violating time, this fact was to be expected from the very
definition of $\zeta_\mathrm{ad}$, see Eq.~(\ref{zetadef}).

One point of interest to be mentioned here is the fact that, if one
restricts attention to adiabatic perturbations, then both variables
$u$ and $v$ exhibit divergences while passing through the
bounce. However, those present in $u$ stem from the NEC violation,
necessary in this case, and can be understood as mere computational
artifacts as they are exactly compensated in the calculation of
$\Phi_\mathrm{ad}$, which thus appears to be the regular variable to
consider in this instance (remember that the divergence in
$\Phi_\mathrm{ad}$ appears only in its second derivative, at order
$k^4$ in the above example). In contrast, $v$ is actually unbounded
also at the bounce itself, as well as at the points for which the
sound velocity vanishes, while $\zeta_\mathrm{ad}$ exhibits a pole
around the NEC transition. 

This result merely means that the comoving hypersurfaces are not
defined at the NEC violating points, which is hardly surprising if one
recalls the definition of these hypersurfaces~\cite{bardeen}: they are
given by the requirement that the total stress-energy tensor satisfy
$T_i^0 =0$. Setting $\delta p_\mathrm{com}$ the pressure perturbation
with respect to these hypersurfaces, the comoving coordinates are then
completed by the comoving time $t_\mathrm{com}$ related to the cosmic
time $t$ through the relation~\cite{lyth}
\begin{equation}
\dd t = a \dd\eta = \left( 1-\frac{\delta p_\mathrm{com}}{\epsilon +
p}\right) \dd t_\mathrm{com}.\label{comoving}
\end{equation}
It is clear that the divergence observed in the curvature
perturbation can then be interpreted as a bad choice of
coordinates, the transformation to the comoving coordinates being
singular at the NEC transition, and hence forbidden.  It seems
therefore that, at least in this example, propagating $u$, \ie
$\Phi_\mathrm{ad}$, through the bounce and NEC transition makes more
sense than propagating $v$, \ie $\zeta_\mathrm{ad}$.

In the case one tries to match the collapsing phase to the expanding
phase directly, without taking into account the detailed structure of
the bounce, then neither $\Phi$ nor $\zeta$ is actually continuous
through that particular bounce; in fact, it is the combination $\beta
\left( \zeta-\Phi\right )/\Hu$ which passes continuously in this
case~\cite{ppnpn2}. This bears close resemblance with another
situation in which general relativity alone is used to perform the
bounce~\cite{turok,jmpp}; these results provide new example for the
general framework set in Ref.~\cite{cdc}.

\section{Conclusions}\label{sec:con}

In this paper, we have examined with greater details one
particular case previously discussed in Ref.~\cite{nobounce} in
which adiabatic perturbations in the hydrodynamical framework
present divergences in its second derivative around the NEC
transition, see Eq.~(\ref{divnec}), even though, as inspection of
Eqs.~(\ref{ap3}) and (\ref{ap4}) reveals, the full Bardeen
potential and all its derivatives are perfectly regular.
Eqs.~(\ref{ap3}) and (\ref{ap4}) constitute a set of equations
which describe the Bardeen potential, hence the scalar part of
the metric fluctuations, completely. An alternative set of
equations is the one based on the definition of entropy
fluctuations (essential for the definition of an adiabatic
perturbation) given in Eq.~(\ref{defmodes}), namely,
Eqs.~(\ref{eqS27}) and (\ref{bard2}). These equations present
divergent coefficients essentially because Eq.~(\ref{defmodes})
contains $\cs$, which has a pole at the NEC transition, see
Eq.~(\ref{csz}).  This fact suggests that the system
(\ref{eqS27})--(\ref{bard2}) may not be appropriate to describe
perturbations around the NEC transition. However, by combining
these two equations, we have obtained a completely regular fourth
order equation for the full Bardeen potential, where no
divergences in their solutions appear at any order, even though
the equations from which it was obtained presented coefficients
with singular points at the NEC transition. Hence, the hydrodynamical
treatment of the example based on Eqs.~(\ref{eqS27}) and
(\ref{bard2}) is completely equivalent to the one in terms of
fields worked out in Ref.~\cite{ppnpn2}, where the same fourth
order equation was obtained.

Nevertheless, inspection of Eq.~(\ref{eqS27}) shows that entropy
fluctuations cannot be neglected, even for arbitrarily small but
non vanishing values of the wavelength $k$, at the NEC transition
time as long as its source term diverges there. Hence, adiabatic
perturbations cannot be defined at this point and the divergence
detected in Ref.~\cite{nobounce} is not physically meaningful.
That may be understood in much the same way as undamped
resonances or shock waves: if one neglects the friction terms
thanks to which the energy of the modes can be evacuated away,
which is what we do by considering only adiabatic perturbation,
one introduces artificial discontinuities (in the case of shocks)
or even divergences.  The adiabatic perturbation approximation,
which does not take into account possible exchanges between the
various fluids involved, and therefore does not count correctly
the relevant degrees of freedom, can only make sense if one of the
fluids clearly dominates over the others, or if there is only one
fluid with complicated equation of state, as was considered in
Ref.~\cite{nobounce}.

The case studied in section~\ref{sec:pertspec} is but an example
for which the scale factor near the bounce can be expanded as a
symmetric power series in the conformal time $\eta$, namely as
$a(\eta)=a_0 + b\eta ^2 + e\eta ^4 + \cdots$, where $a_0$, $b$
and $e$ are constants (it can be shown that the coefficients up
to the fourth derivative of the scale factor are necessary to
describe the passing of the perturbations through the
bounce~\cite{jmpp}). We have shown in section~\ref{sec:back} that
this is the only possible behavior of the scale factor in a model
where the energy density can be written as a function of $a$.
This situation comprehends the case of many non interacting
fluids with constant equations of state, where we have proven
that [see Eq.~(\ref{apg})] the full Bardeen potential and all its
derivatives are completely regular at all times.

We have also shown in section~\ref{sec:genpert} that for models with
other behaviors near the bounce, the conclusions of
Ref.~\cite{nobounce} hold correct, namely, that, besides divergences
at the NEC transition, there are divergences in the adiabatic
perturbation in the bounce itself.  As the matter content of these
types of bounce cannot be modeled with many non interacting fluids
with constant equations of state, one cannot prove in the way we did
that the full Bardeen potential is finite in such cases, if it is. For
{\it case (1)} of section~\ref{sec:genpert}, as the divergence in the
adiabatic Bardeen potential comes in zero order in $k$, we can only
conjecture that they are genuine divergences of the full Bardeen
potential. For the other cases, as they come in at least second order
in $k$, it may be that a process analogous to the one described in
section~\ref{sec:pertspec} occurs, and the adiabatic treatment alone
may not be meaningful also at the bounce itself.  In order to see if
this is the case, specific examples of this type must be
constructed. In fact, nonsymmetric bounces may be physically more
interesting than the symmetric ones. Indeed, some string/brane
motivated models~\cite{ekp} have been proposed in the literature,
which, although mathematically inconsistent for various
reasons~\cite{noekp}, suggest that the turning point, \ie the bounce
itself, could be the time at which entropy is produced, in the form of
radiation say, not in a smooth way. Seen from a four dimensional point
of view, such an entropy production can be modeled by a nonsymmetric
bouncing scenario, since the total energy density is then no longer a
function of the scale factor. It is clear however that, according to
the present analysis, the fate of linear perturbation theory in
general relativity bouncing models cannot be guaranteed by examining
adiabatic perturbations alone. Also, in the case linear perturbation
theory really breaks down, it is possible that such models do have
important second order contributions, which means possibly strong
backreaction effects. All these questions deserves further
investigation.

\acknowledgments

We acknowledge CNRS and CNPq for financial support. We also would like
to thank David Langlois, J\'er\^ome Martin, Raymond Schutz and
Jean-Philippe Uzan, and the group of ``Pequeno Semin\'ario'' for
various enlightening discussions. Special thanks are due to Robert
Brandenberger for various comments concerning ``entropy'' modes and
the power series expansion of the scale factor that led to
considerable improvement of this work. PP would like to thank the
Abdus Salam ICTP (Trieste, Italy) for hospitality during the latest
stages of this work. NPN would like to thank Jos\'e Martins
Salim for very useful discussions.

\end{document}